\documentclass{elsart}
\usepackage{epsfig}

\begin{document}

\begin{frontmatter}

\title{\Large\bf\boldmath Study of $J/\psi \to \omega K^+K^-$}

\date{\today}

\maketitle

\begin{center}
\begin{small}
\vspace{0.2cm}

M.~Ablikim$^{1}$, J.~Z.~Bai$^{1}$, Y.~Ban$^{10}$,
J.~G.~Bian$^{1}$, D.~V.~Bugg$^{19}$, X.~Cai$^{1}$, J.~F.~Chang$^{1}$,
H.~F.~Chen$^{16}$, H.~S.~Chen$^{1}$, H.~X.~Chen$^{1}$
% ...
\\(BES Collaboration)\\

\vspace{0.2cm}
\label{att}
$^1$ Institute of High Energy Physics, Beijing 100039, People's Republic of China\\
$^2$ China Center for Advanced Science and Technology(CCAST), Beijing 100080,
People's Republic of China\\
$^3$ Guangxi Normal University, Guilin 541004, People's Republic of China\\
$^4$ Henan Normal University, Xinxiang 453002, People's Republic of China\\
$^5$ Huazhong Normal University, Wuhan 430079, People's Republic of China\\
$^6$ Hunan University, Changsha 410082, People's Republic of China\\
$^7$ Liaoning University, Shenyang 110036, People's Republic of China\\
$^8$ Nanjing Normal University, Nanjing 210097, People's Republic of China\\
$^9$ Nankai University, Tianjin 300071, People's Republic of China\\
$^{10}$ Peking University, Beijing 100871, People's Republic of China\\
$^{11}$ Shandong University, Jinan 250100, People's Republic of China\\
$^{12}$ Shanghai Jiaotong University, Shanghai 200030, People's Republic of China\\
$^{13}$ Sichuan University, Chengdu 610064, People's Republic of China\\
$^{14}$ Tsinghua University, Beijing 100084, People's Republic of China\\
$^{15}$ University of Hawaii, Honolulu, Hawaii 96822, USA\\
$^{16}$ University of Science and Technology of China, Hefei 230026, People's Republic of China\\
$^{17}$ Wuhan University, Wuhan 430072, People's Republic of China\\
$^{18}$ Zhejiang University, Hangzhou 310028, People's Republic of China\\
$^{19}$ Queen Mary, University of London, London E1 4NS, UK \\
\vspace{0.4cm}
$^{\ast}$ Current address: University of Michigan, Ann Arbor,
Michigan, 48109, USA \\
$^{\dagger}$ Current address: Purdue University, West Lafayette, Indiana 47907, USA.
\end{small}
\end{center}

\normalsize

\begin{abstract}
New data are presented on $J/\psi \to \omega K^+K^-$
from a sample of 58M $J/\psi$ events in the
upgraded BES\,II detector at the BEPC.
There is a conspicuous signal for $f_0(1710) \to K^+K^-$
and a peak at higher mass which may be fitted with
$f_2(2150) \to K\bar K$.
From a combined analysis with $\omega \pi ^+ \pi ^-$ data,
the branching ratio $BR( f_0(1710)\to\pi\pi )/BR( f_0(1710)
\to K\bar K)$ is $< 0.11$ at the $95\%$ confidence level.

\vspace{5mm}

\end{abstract}

\end{frontmatter}
\clearpage

In a recent publication, we have presented new data on $J/\psi \to
\omega \pi ^+ \pi ^-$ [1] from a sample of 58M $J/\psi$ events
taken in the Beijing Spectrometer (BES) detector at the Beijing
Electron Positron Collider.
Here we report data on $J/\psi \to \omega K^+K^-$.
Earlier data on this channel with lower statistics have been
published by Mark I [2], DM2 [3] and Mark III [4].

The BES II detector is a large solid-angle magnetic spectrometer that
is described in detail in Ref. [5].
Charged particles are measured in a vertex chamber and
Main Drift Chamber (MDC); these are surrounded by a solenoidal
magnet providing a nearly uniform field of 0.4T.
Photons are detected in a Barrel Shower Counter (BSC) made of gas
proportional tubes interleaved with 12 radiation lengths of lead sheets.
A time-of-flight (TOF) hodoscope immediately outside the MDC provides
separation between pions, kaons, and protons.  The time resolution of
the TOF measurement is 180 ps.
Further separation is obtained using $dE/dx$ in the MDC.

The point of closest approach of a charged track to the beam is
required to be within 2 cm of the beam axis and within 20 cm of the
centre of the interaction region along the beam axis. Both photons are
required to be isolated from charged tracks by demanding an angle
$>8^{\circ}$ to the nearest charged track. Any photon with an energy
deposit $< 30$ MeV in the shower counter is rejected. All
particles are required to lie well within the acceptance of the
detector, with charged tracks having laboratory polar angles $\theta$
satisfying $|\cos \theta | < 0.84$ and with transverse momenta
$> 60$ MeV/c.

The $\omega$ is observed decaying to $\pi ^+ \pi ^- \pi ^0$, so
events are selected initially by demanding two photons and four
charged tracks with total charge zero.
If there are more than two photons, all are tried; an extra photon
can arise from interactions of charged particles in the detector.
Kaons can be identified up to momenta of 800 MeV/c by TOF and $dE/dx$
measurements.  The two slowest particles always have energies $<800$
MeV.
The first step is to identify one kaon and one pion
using TOF and $dE/dx$. The other two tracks often have momenta
too high to be identified by TOF and $dE/dx$, so a four-constraint
kinematic fit is made for the $K^+K^-\pi ^+ \pi ^-
\gamma \gamma$ hypothesis.
The kinematic fit
requires $\chi^2 (K^+K^- \pi ^+\pi ^-\gamma \gamma ) <40$.

The $\pi ^0$ is selected by requiring
$|M_{\gamma \gamma } - M_{\pi ^0}| < 0.020$
GeV/c$^2$; the $\pi ^0$ mass resolution is $\sim 15$ MeV/c$^2$.
The resulting $\pi ^+\pi ^- \pi ^0$ mass distribution is
shown in Fig. 1.
The $\omega$ signal is then selected requiring
$|M_{\pi ^+ \pi ^- \pi ^0} - M_{\omega }| \le 40$ MeV/c$^2$.
The background is fitted by a second order polynomial in
$M(\pi ^+ \pi ^- \pi ^0)$.
A background of $(22.9\pm 2.0)$\% is estimated from $\omega$
sidebands, defined by
$80 \le |M_{\pi ^+ \pi ^- \pi ^0} - M_{\omega }| \le 160$ MeV/c$^2$;
the error allows for small variations when the location and width
of the sidebins are changed.

For a given $\omega$ momentum, the mass of the
accompanying $K\bar K$ pair is unique.
The decay angles of $\pi \pi$ and $K\bar K$ in the lab frame are very
different except near 0 or $180^\circ $.
There, the backward $\pi$ or $K$ differ strongly in momentum
and are easily distinguished by momentum, TOF, and $dE/dx$.
As a result, there is a clean separation between $\omega \pi^+ \pi^-$
and $\omega K^+K^-$.

Most background originates from $K^+K^-\pi ^+ \pi ^- \pi ^0$.
The other sources of background are $K_S^0$ in
final states $K^0_SK^{\pm }\pi ^{\mp }\pi^0$ and
$K^0_SK^{\pm }\pi ^{\mp }\gamma$.
Most $K_S$ events are rejected as follows.
If $\chi^2 (K^0_SK^{\pm }\pi ^{\mp }\gamma  \gamma) <
    \chi^2 (K^+K^- \pi ^+\pi ^-    \gamma  \gamma )$ or
   $\chi^2 (K^0_SK^{\pm }\pi ^{\mp }\gamma ) <
    \chi^2 (K^+K^- \pi ^+\pi ^-    \gamma  \gamma )$,
events are discarded if any $K\pi \pi \pi$ combination has
$M(\pi ^+ \pi ^-)$ in the
interval $497 \pm 25$ MeV/$c^2$ and $r_{xy} > 3$ mm;
here $r_{xy}$ is the
distance from the beam axis to the $\pi ^+\pi ^-$ vertex.
This avoids rejecting too many signal events; surviving $K_S$
background is too small to be visible.
The beam spot has a $\sigma_x$ of 0.6 mm, and the resolution of
the second vertex is 1.2 mm in $xy$.
After the background subtraction,
there are 3438 signal events.
From the Monte Carlo simulation, the
average detection efficiency is 4.0\%.

%Fig. 1
\begin{figure}[htbp]
\begin{center}
\epsfig{file=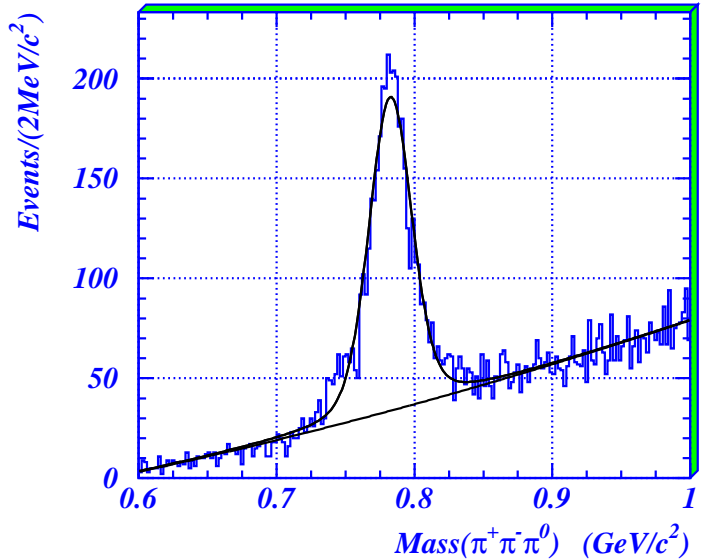,width=10.0cm}
\caption{The $\omega$ peak from the selection described in the text;
background is estimated from the lower curve.}
\end{center}
\end{figure}

Fig. 2(a) shows the experimental Dalitz plot, and Figs. 2(c) and (d)
show projections on to masses of $K^+K^-$ and $\omega K$;
the shaded area indicates background events from the sideband
estimation.

%FIG 2.
\begin{figure}[htbp]
\begin{center}
\epsfig{file=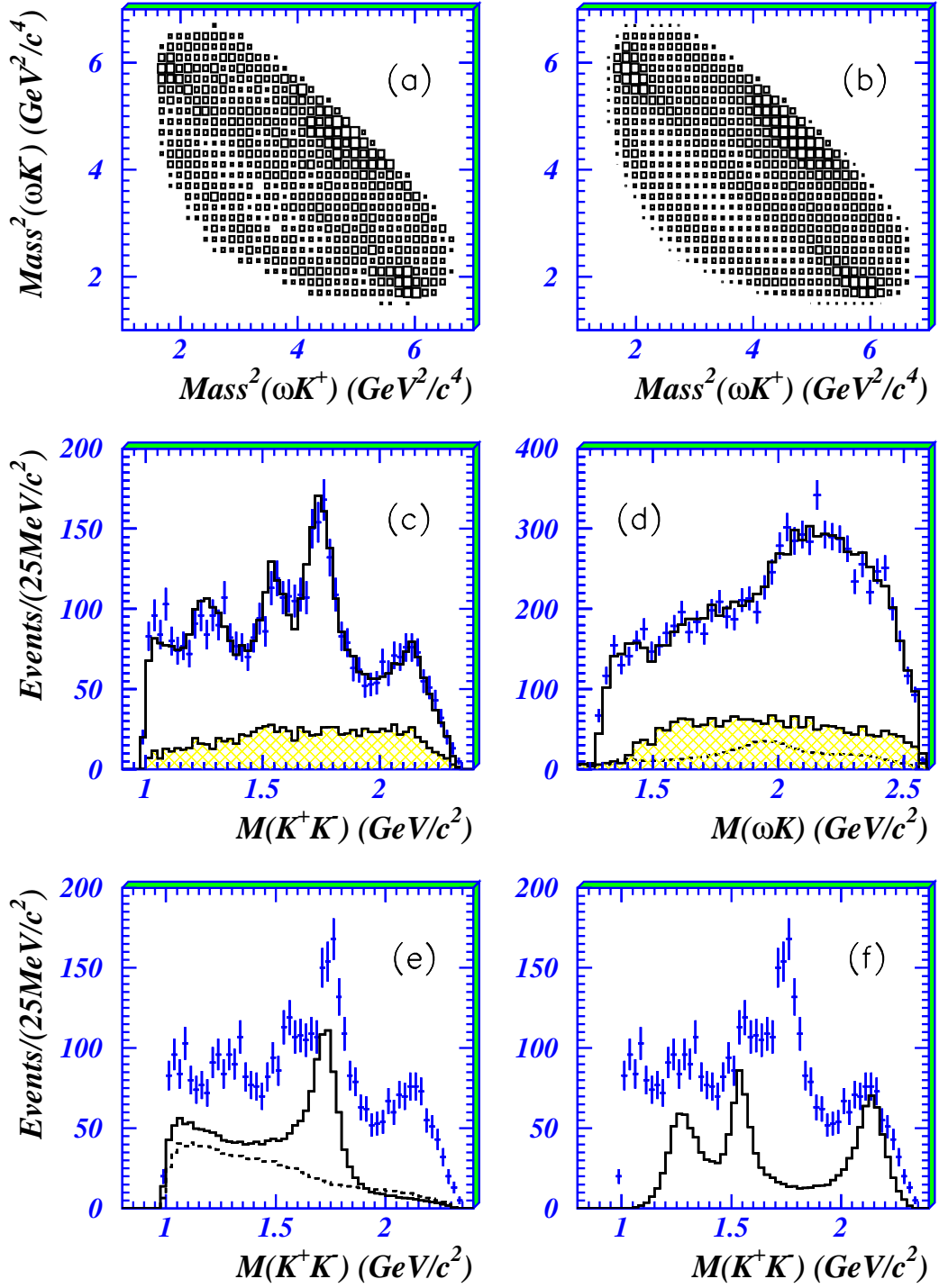,width=14.0cm}
\caption[]{(a) and (b): Measured and fitted Dalitz plots for
$\omega K^+K^-$; (c) and (d) are projections on to $K^+K^-$
and $\omega K$ mass.
In the latter, histograms show the maximum likelihood fit;
the shaded region indicates the background estimated from
sidebins; the dashed curve in (d) shows the magnitude of the
$K_1(1400)$ contribution and a $K\omega$ contribution
at 1945 MeV/c$^2$;
(e) and (f) show mass projections of $f_0$ and $f_2$ contributions
to $K^+K^-$.
The dashed curve of (e) shows the $\sigma \to K^+K^-$ S-wave
contribution.}
\end{center}
\label{figure2}
\end{figure}

The channels fitted to the data are:
\begin {eqnarray}
J/\psi &\to & \omega \sigma \nonumber \\
      &\to & \omega f_0(980) \nonumber \\
       &\to & \omega f_0(1710) \nonumber \\
       &\to & \omega f_2(1270) \nonumber \\
       &\to & \omega f_2'(1525) ~{\rm or}~\omega f_2(1565) \nonumber \\
       &\to & \omega f_2(2150)\nonumber \\
       &\to & K_1(1400)K \nonumber \\
       &\to & K_1(1950)K .\nonumber
\nonumber
\end {eqnarray}
Amplitudes are fitted to relativistic tensor expressions
documented in Ref. [6].
For spin 0 in $K\bar K$, two transitions from $J/\psi$ are allowed
with orbital angular momenta $\ell = 0$ and 2 in the production
process. For spin 2, there are five amplitudes: one with $\ell = 0$,
three with $\ell = 2$ and one with $\ell = 4$.
In fitting these, Blatt-Weisskopf centrifugal barrier factors are
included with a radius of 0.8 fm, though results are insensitive to
this choice.
In the amplitude analysis, information from the $\omega \to \pi ^+ \pi
^- \pi ^0$ decay is included in the tensor expressions.

The polarisation vector of the $\omega$ lies along the normal
to its decay plane.
The correlation between this polarisation vector, the production
plane,  and the decay of the $f_J$ to $K^+K^-$ is sensitive to
the spin of $f_J$ and also to the helicity amplitudes for its
production.
This correlation cannot readily be displayed, since it depends
on five angles; however, tests with different $J^P$ demonstrate the
sensitivity to quantum numbers.

Fig. 2(b) shows the Dalitz plot from the log likelihood fit
described below. Histograms on Figs. 2(e) and (f) show
projections of $f_0$ and $f_2$ contributions to this fit.

The $\omega \pi ^+ \pi ^-$ data of Ref. [1] determine all helicity
amplitudes for production of $f_2(1270)$ well.
In fitting present data, the relative magnitudes of these amplitudes
are fixed to values from $\omega \pi \pi$.
Contributions from $f_0(980)$ are likewise fixed from the
signal observed in $\omega \pi ^+ \pi ^-$;  its
branching ratio $K\bar K/\pi \pi$ is taken from the
Flatt\' e formula fitted to $J/\psi \to \phi \pi ^+ \pi ^-$
and $\phi K^+K^-$ [7], where there are conspicuous $f_0(980)$
signals. Phases for $f_2(1270)$ and $f_0(980)$ amplitudes are fitted
freely, since they arise from multiple scattering, which is different
in $K\bar K$ and $\pi \pi$ final states.

For other components, there is a general problem in isolating
$f_0$ from possible $f_2$ for two reasons.
Firstly, five $2^+$ amplitudes can simulate two $0^+$ amplitudes
closely;
amplitudes with $J^P = 2^+$ may be
identified if they give rise to decay angular distributions which are
non-isotropic.
Secondly, fitted $2^+$ amplitudes can fluctuate for
angles outside the acceptance.
For high $K^+K^-$ mass above 2 GeV$/c^2$, this latter problem is
somewhat reduced, because the $\ell = 4$ amplitude is suppressed by the
strong centrifugal barrier for production.

We use $\sigma$ to denote  a broad $K^+K^-$ S-wave
contribution.
We find that it peaks towards the lower $K\bar K$ masses as shown by
the dashed curve of Fig. 2(e).
However, the dependence on mass above
1 GeV is somewhat uncertain.
Many alternative fits have been tried with
similar results.
A component peaking towards threshold is
required; without it, the fit to the $K\bar K$ mass distribution of
Fig. 2(c) is bad.
We have therefore tried parametrisations
using the $\sigma$ pole of Ref. [1], and a coupling
constant of the form $G_1 + G_2 s$ or $G_1 +G_2/s$.
The optimum fit requires a slightly more rapid fall with $s$
than the $\sigma$ pole, in order to fit four points at
the lowest $K\bar K$ masses.
However, we regard this as unphysical and therefore eventually
choose to use the $\sigma $ pole of Ref. [1] unchanged, with $G_2 = 0$.
%At the lowest mass, the fit then lies marginally below
%the data points.
%%One might think that the difference could be explained by
%increasing the $f_0(980)$ contribution.
%That is not the case, since its
%magnitude is limited by the $\omega \pi ^+ \pi ^-$ data
%of Ref. [1].
Note that there is a substantial constructive interference
in present data between $f_0(980)$ and $\sigma$ amplitudes at
masses close to threshold.

A dominant feature is $f_0(1710)$;
the present data are consistent with earlier studies which
identify $J=0$ [8,9]. They are also consistent with the
absence of any significant $J = 2$ contribution.
The fitted $f_0(1710)$ optimises at $M = 1738 \pm 30$ MeV$/c^2$,
$\Gamma = 125 \pm 20$ MeV$/c^2$. The error in the mass is
mostly systematic, and arises from
uncertainty in the $\sigma$ amplitude with which $f_0(1710)$
interferes; the error in $\Gamma$ is mostly statistical, but
includes allowance for interference with the remaining $0^+$ amplitude.
Earlier BES \,II data on $J/\psi \to \gamma K^+K^-$ and $\gamma
K^0_SK^0_S$ gave $M=1740 \pm 4({\rm stat}) ^{+10}_{-25}({\rm syst})$ MeV$/c^2$
and $\Gamma = 166 ^{+5}_{-8}~ ^{+15}_{-10}$ MeV$/c^2$ [8].

A fit to the 1738 MeV/c$^2$ peak with spin 2 uses five amplitudes and gives
log likelihood worse than spin 0 by only 15; the fit is shown in
Fig. 3.
However, the fit with spin 0 uses only two production
amplitudes with $\ell = 0$ and 2.
The fit with spin 0 requires an $\ell = 0$ amplitude
which is completely dominant over $\ell = 2$.
However, for spin 2  the $\ell = 2$ amplitudes dominate
over $\ell = 0$.
The phase space available in the process $J/\psi \to \omega
f_J(1710)$ is rather limited, and the $\ell = 2$ and 4 centrifugal
barriers for the production process should suppress
those amplitudes strongly.
If  the $\ell = 2$ and 4 amplitudes are removed, spin 0 gives a fit
better than spin 2 by 90 in log likelihood.

This pattern of behaviour is symptomatic of what is required for
spin 2 to simulate spin 0.
The spin 2 amplitude with $\ell = 0$ has a unique dependence on
angles; it contains a distinctive
term $3\cos ^2 \alpha _K - 1$, where
$\alpha _K$ is the decay angle of the $K^+$ in the
resonance rest frame, with respect to the direction of the
recoil $\omega$. Simulation of spin 0 requires large $J = 2$
$\ell = 2$ and 4 amplitudes to produce compensating terms in
$\sin ^2\alpha _K$. Although this is suspicious, the $J = 2$ possibility
cannot be ruled out from present data.

%Fig. 3
\begin{figure}[htbp]
\begin {center}
\epsfig{file=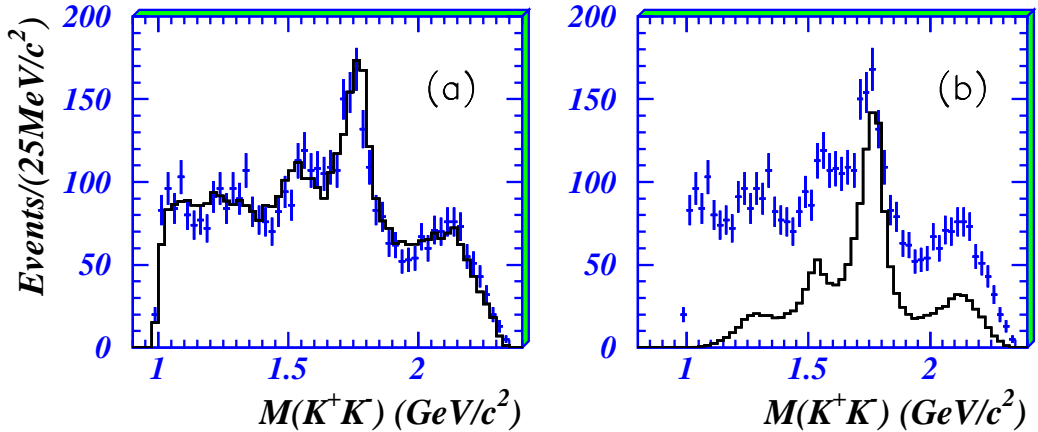,width=14.0cm}
\caption{(a) The projection on to $M(K^+K^-)$ from an alternative fit
using $f_2(1710)$, (b) the contribution from $J^P = 2^+$.}
\end{center}
\end{figure}

We discuss next the branching ratio of $f_0(1710)$ between
$K\bar K$ and $\pi \pi$, using information from $J/\psi \to
\omega \pi ^+ \pi ^-$ [1], where statistics of $\sim 40 K$
events are available.
In those data, there is no definite  evidence for the presence
of $f_0(1710)$ ; if its mass is scanned, there is no
optimum around 1710 MeV/c$^2$, and the fitted $f_0(1710)$ is only $0.43 \pm
0.21$\% of $\omega \pi^+\pi^-$.
In the $\omega K^+K^-$ data presented here,
the $f_0(1710)$ intensity is $(38 \pm 6)\%$ of the data within
the same acceptance as for $\omega \pi ^+ \pi ^-$; the error is
almost entirely systematic, and covers all alternative
parametrisations of the $\sigma$ amplitude and
removing the $K_1(1400)$.
The branching fraction for $J/\psi \to \omega f_0(1710)$,
$f_0(1710)\to K^+K^-$ is $(6.6 \pm 1.3) \times 10^{-4}$.
We find at the 95\% confidence level
\begin {equation} \frac {BR(f_0(1710) \to \pi \pi )}
{BR(f_0(1710) \to K\bar K )} < 0.11,
\end {equation}
where all charge states for decay are taken into account.

One caveat is necessary.
In our study of $J/\psi \to \phi \pi ^+\pi ^-$ and
$\phi K^+K^-$ [7], definite evidence is found for an
$f_0(1770)$, distinct from $f_0(1710)$ and decaying
to $\pi \pi$ (and possibly weakly to $K\bar K$).
There is a remote possibility that $f_0(1710)$ and $f_0(1770)$ are
both present in $\omega \pi \pi$ data but cancel by destructive
interference.
Such a cancellation would require that they have the same magnitudes
but opposite phases.
Even then, the cancellation is incomplete, because they have different
masses and widths.
Allowing for this possible cancellation, the upper limit of the
branching ratio given in eqn. (1) could increase to 0.16 if the
magnitudes happen to be equal, which is unlikely.
%probability of such a chance cancellation is unpredictable.

The peak in Fig. 2(c) at $\sim 1550$ MeV$/c^2$ may be fitted with either
$f_2'(1525)$ or $f_2(1565)$, or both.
Spin 2 is required by non-isotropic decay angular
distributions; a fit with an $f_0$ with the same mass and width gives
a worse log likelihood by 64. Also no $f_0(1500)$ is visible in the
$\omega \pi ^+\pi ^-$ data of Ref. [1].
If the peak is fitted with $f_2'(1525)$, the branching fraction is close
to that for $f_2(1270) \to K\bar K$. However, because of interferences
between helicity amplitudes, the branching fraction could be a factor
2 larger or smaller.
If the peak is fitted with $f_2(1565)$, the
branching fraction is similar to that of $f_2(1565) $ in $\omega \pi
\pi$ data, but again could be a factor 2 larger or smaller.
The fit shown in Fig. 2 uses $f_2'(1525)$.
The branching ratio of $f_2(1270)$
between $K\bar K$ and $\pi \pi$ is $(5.2 \pm 2.5)\%$, consistent with
the range of values quoted by the Particle Data Group [9]; again the
error arises from flexibility in interferences between helicity
amplitudes.

There is a further feature at $\sim 2150$ MeV/c$^2$ in the $K^+K^-$
mass spectrum.
Some spin $\ge 2$ component is required by non-isotropic
decay angular distributions.
An optimum fit to present data may be achieved with a mass of
$2150 \pm 20$ MeV$/c^2$ and a width $\Gamma = 150 \pm 30$ MeV$/c^2$;
these values are within a few MeV/c$^2$ of PDG values.
Errors are mostly statistical but also cover changes when the
small amplitudes are omitted from the fit.
The data do not rule out the possibility of spin 4, but
the fit is consistent with the known $f_2(2150)$ [9].

%Fig. 4
\begin{figure}[htbp]
\begin{center}
\epsfig{file=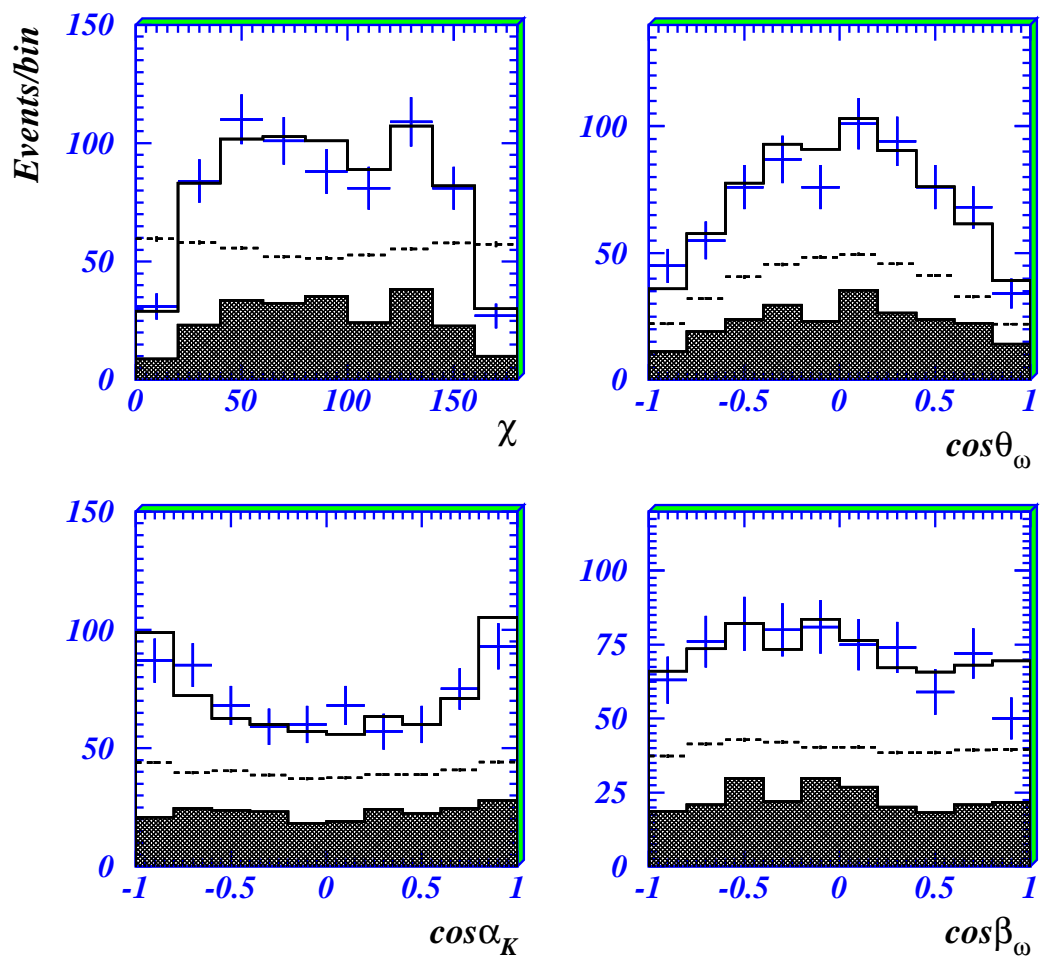,width=14.0cm}
\caption[]{Angular distributions for $M_{KK} > 2000$ MeV$/c^2$
for angles $\chi$, $\theta _\omega$, $\alpha _K$ and $\beta _\omega$
defined in the text; histograms show the fit and the lower
shaded histograms the background, taken from sidebands.
The dashed histograms show the acceptance.}
\end{center}
\label{figure22}
\end{figure}

Fig. 4 shows distributions for four angles after selecting
$M_{KK} > 2000$ MeV$/c^2$.
The angle $\chi$ is the angle between the
decay plane of $\omega \to \pi^+\pi^-\pi^0$ and
the decay plane $X \to K \bar K$;
$\theta _\omega$ is the production angle
of the $\omega$ in the $J/\psi$ rest frame.
The angle $\alpha _K$ is the decay angle of the
$K$ in the rest frame of $X$, taken with
respect to the direction of the recoil $\omega$;
$\beta _\omega$ is the angle between the normal to the
$\omega$ decay plane and the beam direction.
The distribution for $\cos \alpha _K$ is distinctly
non-isotropic, although after integrating over all but
one of the angles, much of the spin information is lost;
the full amplitude analysis is much more reliable than
projections on to individual angles.
The dashed curves illustrate the acceptance. The
shaded histograms at the bottom of each panel
show background, which is taken from sidebins.

A marginal improvement of 21 in log likelihood may be obtained by
adding $f_0(2100) \to K^+K^-$.
However, this is not sufficient to be sure of its presence, so it is
omitted.

%Fig. 5
\begin{figure}[htbp]
\begin{center}
\epsfig{file=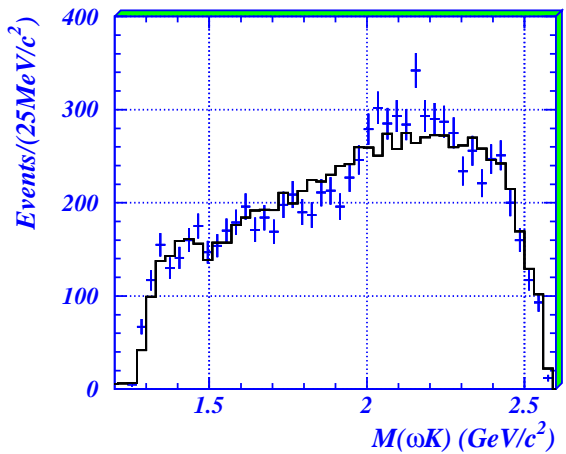,width=8.0cm}
\caption{ The projection on to $M(\omega K^\pm)$ omitting
the $\omega K$ signal at 1945 MeV$/c^2$.}
\end{center}
\end{figure}

   The $\omega K$ mass distribution is fitted poorly unless a
component decaying to $\omega K$ is included with $M \sim 1945$
MeV$/c^2$, $\Gamma \sim 270 $ MeV$/c^2$.
Fig. 5 shows the poor fit without this additional
amplitude; no background appears capable of explaining this
effect.
The optimum fit is obtained with orbital angular momentum
$\ell = 0$ in the $\omega K$ system, i.e. $J^P = 1^+$;
this improves log likelihood by 113, and contributes 9.4\% of all
events.
The observed isotropic decay can be fitted not only by
$J^P = 1^+$, but through conspiracy between several production
amplitudes for $J^P = 2^-$ and $0^-$.
The known $K_2(1820)$ with $J^P = 2^-$ and
$K(1830)$ with $J^P = 0^-$ [9] do not alone give an adequate
fit but may make some contribution.
Our conclusion is that some $\omega K$ contribution is needed
in this mass range, but cannot be identified cleanly and could
be a superposition of more than one resonance with $J^P = 1^+$,
$0^-$ and $2^-$. Conclusions about $f_0$ and $f_2$ components
are insensitive to this ambiguity.
At lower masses, inclusion of $K_1(1400) \to \omega K$ also gives a
significant improvement of 81 in log likelihood.

%%%%%%%%%%%%%%%%%%%%%%%%%%%%%%%%%%%%%%%%%%%%%%%%%%%%%%%%%%%%%%%%%%%%

In summary, the main features of the data are peaks which may be
attributed to $f_0(1710)$, $f_2(2150)$, $f_2(1270)$ and
either $f_2'(1525)$ or $f_2(1565)$.
An upper limit of 0.11 is set on the
ratio $BR[f_0(1710) \to \pi \pi]/BR[f_0(1710) \to K\bar K]$.
This upper limit could rise to 0.16 if there is a fortuitous
cancellation of $f_0(1710)$ and $f_0(1770)$ in $\omega \pi \pi$
data in both magnitude and phase.

{\vspace{0.8cm}
  The BES collaboration thanks the staff of BEPC for their hard efforts.
This work is supported in part by the National Natural Science Foundation
of China under contracts Nos. 19991480,10225524,10225525, the Chinese
Academy of Sciences under contract No. KJ 95T-03, the 100 Talents Program of CAS
under Contract Nos. U-11, U-24, U-25, and the Knowledge Innovation Project
of CAS under Contract Nos. U-602, U-34(IHEP); by the National Natural Science
Foundation of China under Contract No.10175060(USTC),
No.10225522 (Tsinghua University); and the U.S. Department
of Energy under Contract No.DE-FG03-94ER40833 (U Hawaii).
We wish to acknowledge financial support from the Royal
Society for collaboration between the BES group and Queen Mary,
London.


\begin {thebibliography}{99}
\bibitem {1} J.Z. Bai {\it {et al.}}, (BES Collaboration),
{\it The $\sigma$ Pole in $J/\psi \to \omega \pi ^+ \pi ^-$},
Phys. Lett. B (to be published) and hep-ex/0406038.
\bibitem {2} G.J. Feldman {\it {et al.}}, Phys. Rev. Lett. {\bf 33C}, 285 (1977).
\bibitem {3} A. Falvard et {\it {et al.}}, Phys. Rev. {\bf D38}, 2706 (1988).
\bibitem {4} L. K\" opke and N. Wermes, Phys. Rep. {\bf 174}, 67 (1989).
\bibitem {5} J.Z. Bai {\it {et al.}}, (BES Collaboration),
Nucl. Instr. Meth. {\bf A458},  627 (2001).
\bibitem {6} B.S. Zou and D.V. Bugg, Euro. Phys. J {\bf A16}, 537 (2003).
\bibitem {7} J.Z. Bai {\it {et al.}}, (BES Collaboration),
{\it Study of $J/\psi \to \phi \pi ^+ \pi ^-$ and $\phi K^+K^-$},
to be submitted to Phys. Lett. B.
\bibitem {8} J.Z. Bai {\it {et al.}}, (BES Collaboration),
Phys. Rev. {\bf D68}, 052003 (2003).
\bibitem {9} S. Eidelman {\it {et al.}}, (Particle Data Group), Phys. Lett.
{\bf B592}, 1 (2004).
\end{thebibliography}
\end{document}